\documentstyle[preprint,aps]{revtex}
\tightenlines
\begin{document}
\draft
\title{Current Density Functional approach to large quantum dots
in intense magnetic fields}
\author{M. Pi$^1$, M. Barranco$^1$, A. Emperador$^1$,
E. Lipparini$^{1}$\cite{perm}, and Ll. Serra$^2$.}
\address{$^1$Departament d'Estructura i Constituents de la Mat\`eria,
Facultat de F\'{\i}sica, \\
Universitat de Barcelona, E-08028 Barcelona, Spain}
\address{$^2$Departament de F\'{\i}sica, Facultat de Ci\`encies,\\
Universitat de les Illes Balears, E-07071 Palma de Mallorca, Spain}
\date{\today}

\maketitle

\begin{abstract}

Within Current Density Functional Theory, we have studied a
quantum dot made of 210 electrons confined in a disk geometry.
The ground state of this large dot exhibits   some features
as a function of the magnetic field ($B$) that
can be attributed in a clear way to the
formation of compressible and incompressible states of the system. The
orbital and spin angular momenta, the total energy, ionization
and electron chemical potentials of the ground state, as well
as the frequencies of far-infrared
edge modes are calculated as a function of $B$, and
compared with available experimental and theoretical results.

\end{abstract}


\narrowtext
\section*{}

\section{Introduction}

The study of the formation of compressible and incompressible states of
quantum dots has attracted an enormous interest in the last few years.
It has been argued \cite{ch90,be90,ch92,fe95}
that for a sufficiently
smooth confining potential, the edge of the dot exhibits to a certain
degree, narrow incompressible regions where
the electronic density is determined by the
filling factor $\nu$ as $\rho = \nu /(2 \pi$$\cal{L}$$^2)$,
where $\cal{L}$=$(\hbar c/e B)^{1/2}$ is the magnetic
length, and compressible regions where $\rho$
may vary with the radial distance $r$.
These states are at the origin of phenomena such  as, e.g.,
"magic" angular momentum quantum numbers, composite droplets and
edge reconstruction \cite{ma93,ch94}. Some evidence for the formation
of these states has been found in the magnetic field ($B$) dependence
of the
addition energies, i.e., the energies to add one more electron to
the dot \cite{mc91,as92}, and more recently, in the oscillations
of the resonance frequencies of edge magnetoplasmon modes of dots
and antidots \cite{bo96}.

Exact diagonalizations of the electronic
Hamiltonian corresponding to dots with a small number of electrons
$N$ ($N\leq8$) \cite{ma90,pf93,ha9?,ya93} have shown
that these systems have strongly correlated ground
states (g.s.), especially at high magnetic fields. Approximation
schemes such as the Constant Interaction model of Ref. \cite{be91},
the Self-Consistent model of Ref. \cite{mc92}, the Hartree-Fock method
of Refs. \cite{ch94,gu95}, or other mean field approximations
\cite{fo94} which have been used to carry out calculations in larger
dots, do not take into account these correlations, so they may fail
in reproducing a number of g.s. features
that appear as a function of $B$ \cite{ya93}.  It is worth to mention that
all these approaches have been only applied to the
description of medium size ($N<$100) dots and relatively small values
of $B$ due to the technical difficulties one has to face otherwise
because of the high density of single particle (s.p.) states.

Recently,  Current Density Functional Theory (CDFT) \cite{vi87}
has been used
to study g.s. properties of small quantum dots \cite{fe94}. The
accuracy of the method has been tested in the case of two and three
electron systems, for which exact calculations are available
\cite{pf93}. It has been shown \cite{fe94} that CDFT reproduces
accurately the exact results, thus providing a convincing evidence
that it is a powerful tool to quantitatively describe the g.s.
properties of quantum dots in magnetic fields. CDFT has also been
used to study the far-infrared edge magnetoplasmon modes in medium
size quantum dots by means of an equation of motion method
\cite{li97}.

In the present work, we extend the CDFT calculations to the case of a
large dot of radius $R$=1600${\rm \AA}$ containing 210 electrons in a
magnetic field ranging from $B$=0 to $\sim$ 15 T. The far-infrared
spectrum
of this dot has been experimentally studied in \cite{de90}. The
large number of electrons makes this dot closer to the 2D conditions
than any other system previously considered in a calculation of this
kind. Moreover, the high $B$'s here attained allow us to study
g.s's. which are associated in a clear way with bulk incompressible
states with filling factors  $\nu=\frac{2}{3}$, and from $\nu$=1
to 10, and to several incompressible states in between.
Such a rich structure
has not been previously disclosed in a single large dot,
where genuine magnetic effects, even at rather small $B$, are not
masked by surface nor finite size effects.

The density functional we are using does not incorporate one of
the key features to describe the occurence in a single dot
of several incompressible
strip regions in the fractionaly quantum Hall
regime, namely cusps in the exchange-correlation scalar potential at
certain filling factors (see for example Refs. \cite{fe95,he95}).
However, it is able to describe the
appearance of bulk regions with a fractionary filling factor.
This is illustrated in the case of the
 $\nu = \frac{2}{3}$ incompressible state.

To fully display these
magnetic field effects, we have performed the calculations at a
very small temperature ($T$), 0.1 K.
The use of such temperatures in the calculations
is a challenge for large dots and intense magnetic fields. Some
results
%for a parabolic confining potential ???????, as well as results
at higher $T$ are also discussed. Finally, we have obtained the
 far-infrared spectrum of the dot.

\section{Current Density Functional description of quantum dots}

Current Density Functional Theory is comprehensively described in
\cite{vi87}, and has been already applied to quantum dots
\cite{fe94,li97}. In the present work, we shall be considering
$N$ electrons moving in the $z$=0 plane where they are
confined by the dot potential $V_+(r)$ with $r=\sqrt{x^2+y^2}$. On
this system it may act a constant magnetic field in the $z$ direction
described by the vector potential $\vec{A}=\frac{1}{2}(-y,x,0)B$.
Introducing the cyclotron frequency $\omega_c=eB/mc$,
we can write
the CDFT total grand potential $\cal{A}$ = $E-T S-\mu N$
adding to the total energy
\begin{eqnarray}
E &=& \frac{1}{2}\int d\vec{r}\, \tau(\vec{r}\,)+
\frac{\omega_c}{2}\int d\vec{r} \,r \, j_p(\vec{r}\,)+
\frac{1}{8} \omega_c^2\int d\vec{r} \,r^2 \,\rho(\vec{r}\,)+
g^*\mu_B B \sum_i f_i \, s_{z_i}
\nonumber
\\
& &
\label{eq1}
\\
&+& \int d\vec{r} \,V_+(\vec{r}\,) \rho(\vec{r}\,)+
\frac{1}{2} \int\int d\vec{r} \,d\vec{r}\,'\,\frac{\rho(\vec{r}\,)
\rho(\vec{r}\,'\,)}{\mid \vec{r}-\vec{r}\,'\mid} +
E_{xc}(\rho,\xi,\vec{\cal V})  \,\, ,
\nonumber
\end{eqnarray}
the temperature times the total entropy S given by:
\begin{equation}
S = -\sum_i \left[ f_i \ln f_i + (1 - f_i)\ln (1-f_i)\right] \, \, ,
\label{eq2}
\end{equation}
and the electronic chemical potential $\mu$ times the electron number
$N=\sum_i f_i$.

To easy the formulae, we have used effective atomic units, defined by
$\hbar=e^2/\epsilon=m=k_B=$1, where $k_B$ is the Boltzmann's constant
and $m=m^* m_e$ is the electron effective mass.
In this system of units, the length unit is the
Bohr radius $a_0$ times $\epsilon/m^*$,
and the energy unit is the Hartree times $m^*/\epsilon^2$,
which we call respectively, $a^*_0$ and $E^*_H$.
$m^*$ is the electron effective mass
in units of the bare one,
which together with a
dielectric constant $\epsilon$ and a gyromagnetic factor $g^*$ are
characteristics of the semiconductor (for example, $g^*=-$0.44,
$\epsilon$=12.4 and $m^*$=0.067 in GaAs), and
$\mu _B=\hbar e/(2 m_e c)$ is the Bohr magneton \cite{note}.
For GaAs we have
$a^*_0 \sim$ 97.94 ${\rm \AA}$ and $E^*_H\sim$11.86 meV.

In Eqs. (\ref{eq1},\ref{eq2}), $s_{z_i}$ is the $z$-component of the
spin, and $f_i$ is the occupation number of the $i$-s.p. level. The
particle $\rho(\vec{r}$\,), kinetic energy  $\tau(\vec{r}$\,) and
paramagnetic current densities  $\vec{j_p}(\vec{r}$\,) are defined in
terms of the occupation numbers and s.p. wave functions
$\phi_i$ as follows:
\begin{equation}
\rho (r)=\sum_i f_i \mid\phi_i(r)\mid^2
\label{eq3}
\end{equation}
\begin{equation}
\tau (r)=\sum_i f_i \mid\nabla\phi_i(r)\mid^2
\label{eq4}
\end{equation}
\begin{equation}
\vec{j_p}(r)=j_p(r)\hat{e}_{\theta} =
-\frac{1}{r}\sum_i f_i l_i \mid\phi_i(r)\mid^2 \hat{e}_{\theta}\,\, ,
\label{eq5}
\end{equation}
where $\hat{e}_{\theta}$ is the azimuthal unit vector. Due to the
axial symmetry, the electronic wave functions can be factorized as
\begin{equation}
\phi(\vec{r}\,)=e^{-\imath l\theta} u_{nl\sigma}(r) \,\, ,
\label{eq6}
\end{equation}
where $\sigma= \pm $1,
$n$=1,2,3..., and $l=0,\pm 1,\pm 2,\pm 3$... These wave functions
are eigenstates of the angular momentum around the $z$-axis with
eigenvalue $-l$. At $B\neq 0$, the s.p. level $i$ is non degenerate,
and in the above expressions we have used the short notation
$i\equiv\{n,l,\sigma\}$.
$V_+(r)$ is the Coulomb potential generated by a jellium disk of
radius $R$ and areal density $\rho_+ = \rho_b \Theta(R-r)$
\cite{br90}, and the $\rho_b$ value has been fixed by charge
neutrality, i.e., $\pi \rho_b R^2 = N_+ = N$. For the $R$=1600${\rm
\AA}$ and $N$=210 system which corresponds to one of the GaAs dots
studied in \cite{de90}, one has $\rho_b$=0.239 $a_0^{*\,-2}$.
The use of a step function to model the positive neutralizing
background, i.e., a sharp confining potential \cite{ch95},
together with a smooth exchange-correlation potential,
implies that no strips of filling factor smaller
than the one at the center of the droplet are expected to
appear at the edge of the dot. That will be confirmed by the
calculations presented in the next section.

The exchange-correlation energy $E_{xc}(\rho,\xi,\vec{\cal V})$ in Eq.
(\ref{eq1}) has been taken from \cite{fe94}. It
 is a functional of the particle density $\rho$, the local
spin polarization $\xi$
\begin{equation}
\xi(\vec{r}\,)=\frac{\rho_{\uparrow}-\rho_{\downarrow}}
{\rho_{\uparrow}+\rho_{\downarrow}}             \,\,  ,
\label{eq7}
\end{equation}
where $\rho_{\uparrow}(\rho_{\downarrow})$ is the particle density of
spin up(down) electrons, and of the local vorticity $\vec{\cal V}(r)$
\begin{equation}
\vec{\cal V}(r)={\cal V}(r)\,\hat{e}_{\theta}=-\frac{c}{e\,r}\,\frac{\partial}
{\partial r} \left( r\, \frac{j_p}{\rho}\right) \hat{e}_z \,\, .
\label{eq8}
\end{equation}
For further reference, we also define the local magnetization
$m(r)$ as $m(r)=\rho(r)\,\xi(r)$.

To obtain the s.p. wave functions $\phi_i(\vec{r}\,)$
 and occupation numbers $f_i$, one has
to minimize  $\tilde{\cal{A}}=\cal{A}$ $- \sum_i \lambda_i \langle
\phi_i \mid \phi_i \rangle$. Minimizing $\tilde{\cal{A}}$
 with respect to the wave functions
one obtains the Kohn-Sham (KS) equations:
\begin{eqnarray}
& & \left[-\frac{1}{2} \left( \frac{\partial^2}{\partial r^2}
+ \frac{1}{r} \frac{\partial}{\partial r} - \frac{l^2}{r^2} \right)
-\frac{\omega_c}{2} l+ \frac{1}{8} \omega_c^2 r^2
+ V_+(r) \right.
\nonumber
\\
& &
\label{eq9}
\\
&+& \left. \int d\vec{r}\,'\,\frac{\rho(\vec{r}\,'\,)}
{\mid \vec{r}-\vec{r}\,'\mid}
-   \frac{e}{c}\, l \, \frac{A_{xc}(r)}{r} + V_{xc\sigma}(r)
+ \frac{1}{2} g^*\mu_B B \sigma \right] u_{n l \sigma} =
\epsilon_{n l \sigma} u_{n l \sigma} \,\, ,
\nonumber
\end{eqnarray}
with  $\epsilon_i \equiv \lambda_i/ f_i$ and where
\begin{equation}
V_{xc\sigma}(r)=
\left.\frac{\delta E_{xc}(\rho,\xi,\vec{\cal V})}
{\delta \rho_{\sigma}} \right\vert_{\rho_{-\sigma},\, \vec{\cal V}} \, - \,
\frac{e}{c} \vec{A}_{xc}(r) \cdot \frac{\vec{j}_p}{\rho} \,\, ,
\label{eq10}
\end{equation}
being $\vec{A}_{xc}$ the exchange-correlation vector potential
\begin{equation}
\frac{e}{c} \vec{A}_{xc}(r) = \frac{e}{c} A_{xc} \hat{e}_{\theta} =
\frac{c}{e\,\rho} \frac{\partial}{\partial r} \left(
\frac{\delta E_{xc}}{\delta {\cal V}} \mid_{\rho,\,\xi}\right)
\hat{e}_{\theta} \,\, .
\label{eq11}
\end{equation}
Minimizing $\tilde{\cal{A}}$ with respect to $f_i$ and making use
of the KS equations, one obtains
the occupation numbers:
\begin{equation}
f_i = \frac{1}{1+e^{(\epsilon_i-\mu)/T}}
\,\, .
\label{eq12}
\end{equation}
The electron chemical potential is fixed from the normalization
condition:
\begin{equation}
N = \sum_i \frac{1}{1+e^{(\epsilon_i-\mu)/T}}
\,\, .
\label{eq13}
\end{equation}

The KS differential equations and the normalization condition Eq.
(\ref{eq13}) have been solved selfconsistently,
without expanding the s.p. wave
functions in a necessarily truncated basis of Fock-Darwin or Landau
orbitals. Our iterative method  works for weak and strong $B$ fields
as well, for which the effective potential is
very different. It has the advantage of avoiding the study of how
the results depend on the size of the basis.

\section{Ground state results}

We have studied the g.s. of the $N$=210 electrons dot at $T$=0.1 K
varying the
magnetic field from 0 to $\sim$15.5 T in order to explore states
with filling factors $\nu=\frac{2}{3}$ and above.
We recall that integer $\nu$'s also represent the number
of occupied Landau bands,
which we label as ($M,\uparrow$) or ($M,\downarrow$),
depending on the spin of the s.p. states in the band.
$M=n+{(\vert l\vert-l)/2}$ is the Landau level index.

The density and single particle energies corresponding to the $B$=0
case are represented in Fig. (\ref{fig1}).
The s.p. energies are distributed
along parabolic-like curves as a function of $l$, each curve being
characterized by a different value of $n$.
The figure shows the well
known fourfold degeneracy of each single particle level, corresponding
to the possibilities $\pm l$, ($\uparrow \downarrow$). It is
remarkable that even for such a large system, the density still
displays many oscillations in the bulk which have to be gradually
washed out as the dot evolves towards the 2D system.

Figure (\ref{fig2}) shows the CDFT s.p. energy spectra for $B$=
10.28, 5.14, 3.43, 2.57, 1.71, and 1.29 T as a function of $n$ and $l$.
 Observe that due to the sign of the linear $l$-term in
Eq.\ (\ref{eq9}), most occupied levels have now a positive $l$ value, i.e.,
a negative orbital angular momentum, and that the
($M,\uparrow$) band lies below the ($M,\downarrow$) one because of
the negative value of $g^*$.
Fig. (\ref{fig3}) shows the corresponding g.s. particle $\rho(r)$
and local magnetization $m(r)$ densities together with the
predicted values of total orbital and spin angular momentum
$(L_z, 2S_z)$.

From these figures, one can see that these states are
the finite-size analog of the 2D
Landau incompressible states with $\nu$=1 to 4, 6 and 8
which, for an electronic density
$\rho_e=\rho_b$=0.239 $a_0^{*\,-2}$, would be precisely
realized at the above $B$ values.  Indeed, the densities in
Fig. (\ref{fig3}) exhibit a step-like shape whose plateau has
a density determined by an integer
filling factor $\nu$ as $\rho=\nu e B/ 2 \pi c$.
To help follow the discussion of the results we present, we have
collected in Table I the values of $B$ and $\cal{L}$ that correspond
to $\nu$=1 to 10.
Notice that even for $\nu$=10, $R$ is much larger than $\cal{L}$, so
that the filling factor $\nu$ can be safely used to describe the
occupation of the Landau bands in the interior of the droplet.

The g.s.\ values of the orbital angular momentum agree well with the law
\begin{equation}
L_z  =   \nu^{-1} \,\frac{N(N-1)}{2}  \,\, ,
\label{eq14}
\end{equation}
and the spin is approximately given by
\begin{equation}
2S_z  =  \left\{ \begin{array}{ll}
0              &    \nu \,\,{\rm even}  \\
\frac{N}{\nu}  &    \nu \,\,{\rm odd}  \,\, ,
                \end{array}
\right.
\label{eq15}
\end{equation}
expressions
that are valid in the big-$N$ limit for the Landau bands. This is
explicitly shown
in Fig. (\ref{fig4}), where we have plotted the calculated orbital and
spin angular momenta as a function of $B$.
The small plateaus along the $L_z(B)$ curve at integer $\nu$, also
visible in the $2S_z(B)$ curve, show the robustness of these states
against changes in $B$, and their tendency to not to change their
($L_z,S_z$) content.

In panel (a) of Fig. (\ref{fig3}) we have plotted the density of
the maximum density droplet (MDD) state\cite{mac93}:
\begin{equation}
\rho_{MDD}(r)=  \frac{1}{2 \pi {\cal L}^2}\,\,
 \sum_{l=0}^{N-1} \,\,
\frac{1}{l!}\,\,\left(\frac{r}{\sqrt{2}\,{\cal L}}\right)^{2\,l}
\,\,e^{-\frac{r^2}{2\,{\cal L}^2}}  \,\, .
\label{eq16}
\end{equation}
One can see that both densities are very similar, the only difference
appears in the edge region. The lump in the CDFT density at the
edge is due to the rather sharp confining potential, on the one
hand, and on the other hand to
the exchange-correlation energy, which is changing
rapidly in the surface region of the droplet, see Eqs.
(\ref{eq10},\ref{eq11}), and Refs. \cite{he95}.

It is also interesting to look at the energy difference between the
($M,\uparrow$) and ($M,\downarrow$)
Landau bands shown for example in Fig. (\ref{fig2}).
It is very small for even $\nu$, which correspond to
g.s.'s having $S_z\simeq$0, because the Zeeman term in Eq. (\ref{eq1})
is also small (0.127 meV at $B$=5 T, for instance).
On the contrary, that separation is sizeable for the $\nu$=1
 and 3 cases, in the 2-4 meV range. These g.s. states have
$S_z\neq$0. One should have in mind
that $\omega_c$, which gives the difference of frequencies
between Landau levels when spin and interaction effects are neglected,
is around 18 meV for $B$=10 T.  $V_{xc\sigma}$
is the origin of the large spin splittings. The importance of
the exchange energy in the spin splitting of the Landau bands
was already pointed out in \cite{gu95}.
The similarity of the g.s. corresponding to $\nu$=1
and the MDD state is due to $V_{xc\sigma}$,  as it prevents the
($1,\uparrow$) and ($1,\downarrow$) Landau bands from being close at
$B$=10.28 T and thus, to contribute both to the building of the g.s.

We have not attempted to disentangle which
part of the splitting we have found is due to the exchange energy,
and which part is due to the electronic correlations.
It would be possible at low $B$ as the exchange-correlation energy
comes from the Tanatar-Ceperly parametrization \cite{ta89}, but not
at high $B$ because of the form of the parametrization given by
Levesque et al \cite{le84}.
Exchange-correlation effects enhanced by the finite size of the
system are also responsible for the differences between the calculated
values of $L_z$, and especially of $S_z$, and the ones given by Eqs.
(\ref{eq14},\ref{eq15}). In the case of
$S_z$, these differences can be sizeable
for 'large' odd filling factors, see Fig. (\ref{fig4}).

Figures (\ref{fig5}) and (\ref{fig6}) show the s.p. energies and
densities for $B$= 1.9, 3, 4, 7, 9, and 11 T. These values originate
compressible g.s.'s in the dot, and a careful look at these figures
sheds light about the appearance of compressible and incompressible
states in the finite system.

When the magnetic field is increased from that corresponding
to  $\nu$=1,  some electrons are promoted
within the (1,$\uparrow$) band to s.p. states with higher $l$ values.
This is seen in panel (f) of Fig. ({\ref{fig5}).
Indeed, it can be realized from that figure that the chemical potential
intersects the s.p. curve at a value of $l$ larger than $210$, and
that at the same time, some low-$l$ s.p. levels are at an energy
slightly above $\mu$. This is precisely the evolution
of the MDD state when $B$ is increased and edge reconstruction occurs
\cite{kl95,fe97}.

A different behavior is shown in the other panels of Fig.
(\ref{fig5}). Consider for instance, the situation displayed in
panel (d). Around 140 electrons are in the (1,$\uparrow$) band,
as may be inferred from the intersection point of that band with the
$\mu$ line. All other electrons are oddly distributed in the
(1,$\downarrow$) band, starting from the s.p. levels with
higher orbital angular momentum, whose energies are
well below the chemical potential line because of the dip around
$l$=120 caused by the exchange-correlation energy term.
Whenever the partially filled Landau band presents a flat region,
the electron chemical potential  has to be the average s.p. energy
of that region. Otherwise, electrons will continue
to fill s.p. levels with high-$l$ values belonging to the previous band.

The corresponding electron density manifests the peculiarities
of the s.p. spectrum in quite a distinct manner:
the oscillations displayed by the density at $B$=7 T as shown in
panel (d) of Fig. (\ref{fig5}), are washed out at $B$=9 T as shown in
panel (e) of the same figure
because for this magnetic field,  the s.p. energies of the
(1,$\downarrow$) band are distributed more evenly around $\mu$.

The evolution of the compressible region between $\nu$=2 and 1
when $B$ increases from 5.14 to 10.28~T deserves further discussion, as
this region extends in a wide $B$ range and it
makes the magnetic effects easier to visualize (a similar
evolution has been found between $\nu$=3 and 2). After a regime
extending up to $\geq$ 7 T in which the droplet densities present an
oscillatory behavior, at $B \sim$ 8 T  the (1,$\downarrow$)
flats out and accordingly, the density oscillations disappear.
The configuration at $B$=8 T having $L_z$= 17 031 and $2S_z$= 124
cannot be identified with the $\nu$=1 incompressible state, since
$L_z$ and $2S_z$ are far from having the appropriated values, neither
other similar g.s. configurations  corresponding to
magnetic fields as high as 10 tesla, see Fig. (\ref{fig4}).
It means that looking only at the densities might be misleading
for identifying integer filling factor states.
The evolution of the $B$=10 T state towards the $\nu$=1 one at 10.28 T
proceeds through an interesting change in the s.p. spectra of the
(1,$\downarrow$) band: it bends upwards crossing the $\mu$ line, so
that
its occupied higher-$l$ states  get now empty. To keep constant the
number of electrons on the one hand, and to keep  $L_z$ increasing
with $B$ on the other hand,  the  (1,$\uparrow$) band extends
its occupied s.p. states up to higher-$l$ states. A further increase
in $B$ 'decouples' $\mu$ from the  (1,$\downarrow$) band, and the
situation is eventually that shown in panel (a) of Fig. (\ref{fig2}).

It is of some importance to figure out what is the quantitative
influence of temperature on the above features of the g.s. To this end,
we have carried out calculations using a moderate $T$, namely 2 K.
For quantum dots as the one studied
here, it appears that the density of states around the
chemical potential is large, and  such a small temperature
already produces
changes in the s.p. spectrum and the electron density as well.
This is illustrated in Fig. (\ref{fig7}), where we have drawn the
s.p. energies and electron densities corresponding to
$B$=4 and 7 T.

A comparison of cold and hot s.p. energies in Fig. (\ref{fig7})
shows that a first thermal effect is that the small wiggles around $\mu$
in the cold spectrum are progressively wiped out when $T$ increases.
In other words,
the low-$l$ states which are unoccupied at $T$=0.1 K (they lay
slightly above the chemical potential), are now
occupied with some probability. This flattens the
electron density in the bulk of the droplet.
A second thermal effect is the deepening of the high-$l$ s.p. levels
as $T$ increases. This is due to the widening of the surface region,
a situation that has been
already met in other finite fermionic systems, see for example Refs.
\cite{br74,br91,we92}.
This thermal effect reduces the amplitude of the density
oscillation at the  edge of the dot.

It is thus quite apparent that to study the incompressible states,
one has to use temperatures in the right range of values. Otherwise,
the thermal average implicit in the definition of one-body densities
will wash out the quantum oscillations of compressible states, making
it more difficult the identification of the incompressible ones.

At $B$= 15.42 T, an incompressible state appears,
which deserves a separate analysis.
From the relationship $\rho=\nu e B/ 2 \pi c$, it would correspond to
a filling factor $\nu=\frac{2}{3}$.
 We predict for this state an angular
momentum $L_z$=32 511, which is very close to the one given by
Eq. (\ref{eq14}), namely 32 917. We recall that the angular momentum
of a droplet in a Laughlin's state \cite{la83}
is given by that
equation only when $\nu^{-1}$ equals 1, 3, 5,...
We argue that this formula is a good approximation
not only for integer filling factors and
those of Laughlin's states, but also for  $\nu=\frac{2}{3}$.
As a matter of fact, Fig. (\ref{fig4}) shows that Eq. (\ref{eq14})
works well even for compressible states.

As we show in panel (c) of Fig. (\ref{fig8}), the electron density
is constant in the bulk and has a bump at the dot edge. This has
also been found in simulation calculations of Laughlin's states
\cite{mi93,fe92}. However,
these simulations have been carried out for few electron dots, and
density oscillations are still visible in the bulk. In the CDFT
calculation here presented, the density is constant in the
bulk which, together with the associated values of $L_z$ and $2S_z$,
makes it undubious to label this dot state
with a filling factor $\nu=\frac{2}{3}$.

Further insight into this state can be obtained splitting its density
as follows:
\begin{equation}
\rho_{\frac{2}{3}}(r) = \,\sum^{N-1}_{i=0} \, f_i \mid \phi_i(r) \mid^2
\, + \, \sum_{i\geq N} \, f_i \mid \phi_i(r) \mid^2\, \equiv \,
\rho_{bulk}(r) \, + \, \rho_{surf}(r) \,\, ,
\label{eq16b}
\end{equation}
with $i\equiv \{1,l,\uparrow\}$.
This produces the result displayed in panel (c) of Fig. (\ref{fig8}),
where $\rho_{bulk}(r)$  is represented by a dashed line, and
$\rho_{surf}(r)$  by a dot-dashed line. The integral over $r$
of the former gives $N_{bulk}\simeq$140, and that of the later,
$N_{surf}\simeq$70. This means that two-thirds of the electron number
in the dot is in the bulk region, and one third in the edge region.
A look at  panel (a) of Fig. (\ref{fig8}) shows that the
occupation number of the s.p.\ levels with $l\leq 210$ is quite
similar, around 2/3. Of course, $f_i$ are thermal
occupation numbers, and to have a fractional value, a finite
but small $T$ has to be used in the calculation (0.05 K in the present
case). However, it is a workable
way CDFT has to accomodate noninteger filling
factors in the bulk of an extended droplet.
Whether the occupation numbers represent a 'simple' thermal  situation
or come from
a genuine correlation effect, depends on whether
correlation effects have been smeared out by an unadequate high
temperature or not. We want to stress that the temperature we have
used in the present calculations (50 or 100 mK) is much lower than
the one at which most experiments on quantum dots have been
carried out.
It is worth to recall that within the zero temperature  KS scheme,
fractionary occupation numbers may appear for s.p.\ $i$-states
such that $\epsilon_i=\mu$ \cite{dr90}. The generalization
of Density Functional Theory to this situation
is called  Ensemble Density
Functional Theory, and has been  used in \cite{he95}
to describe quantum dots in the fractionary quantum Hall effect regime.

It is worth to point out that confining potentials
produced by a jellium disk favor the existence
of incompressible regions in the bulk of the dot, since there is
a tendency in the electron bulk density to have a value around that of
the jellium background in order to screen the Coulomb potential.
This can be seen for example in the
compressible states shown in Fig. (\ref{fig6}).
It is also clear that whenever an incompressible state appears, its
bulk density is that of the neutralizing jellium, see Figs.
(\ref{fig3},\ref{fig8}). However, a physically sound
exchange-correlation energy is at the very origin of the
incompressible states we have found.

To substantiate our case, we have displayed in panels (b) and (d) of
Fig. (\ref{fig8})
the results obtained at $B$=15.42 T when the exchange-correlation
energy of \cite{le84} is dropped from the functional. This supports
that the results shown in panels (a) and (c) of that figure are
not only the trivial consequence of screening  the jellium density,
nor of using an unadequate high temperature,
but a product of the high-$B$ correlations   included
in the density functional. Actually, one can see from panel (d)
that density oscillations are clearly visible even at $T$=2 K.

For integer $\nu$ values, the changes are not
so dramatic, and the Tanatar-Ceperly exchange-correlation energy alone
yields results quite similar to those we have shown for $\nu\geq1$.

The $B$ dependence of the total energy $E$ is shown in the top
panel of Fig.\
(\ref{fig9}). Local minima in $E$ are clearly visible at or near
even $\nu$  values corresponding to paramagnetic states of the
droplet having 'zero' total spin momentum.
Local maxima are at or near odd $\nu$ values
which correspond to  ferromagnetic states having large $2S_z$ values,
 but smaller than these given by Eq. (\ref{eq15}). The energy also
presents an inflection point at the ferromagnetic $\nu$=1 state with
$2 S_z$=210. In contradistinction with the situation for
small dots where only one
transition from paramagnetic to ferromagnetic g.s.'s
is observed \cite{pa94}, several smooth transitions
take place below $ B \sim 10$ T for $N$=210.
Also shown in that figure (bottom panel), is the dependence of $E$
with $L_z$, with local minima at the $L_z$ values corresponding to
even filling factors.

The existence of several paramagnetic states plays a crucial role
in keeping the energy of the droplet at around the
value corresponding to $B$=0. $|E|$ decreases
as $S_z$ increases, and viceversa. It is only from the $\nu$=2
state on that $E$ increases monotonously, as $S_z$ does up to
reaching the full polarized value around $\nu$=1.

The electron chemical potential
and the ionization potential $IP= E_{N-1}-E_{N}$ \cite{note2}
are shown in Fig. (\ref{fig10}).
They display a "sawtooth" behavior with $B$, with a fast falling near
integer $\nu$  values. The corresponding $B$ values or
equivalently, the corresponding $L_z$ values, are a kind of
"magic numbers" for which the dot is particularly stable. It is
worth to notice that the large size of the studied dot
is the reason why -$IP$ and $\mu$ are so similar, and why
the oscillations in $IP$ are rather wide,
extending over several tesla and giving
them the possibility of being  experimentally observed
more easily  than in small dots \cite{mc91,as92}.

We have not attempted
a systematic study of the addition energies $E_{ad}$, but
have obtained them for the
$B$=4 and 7 T cases. To this end, we have calculated $E(209)$ keeping
constant $\rho_b$ and adjusting $R$ so as to have $N_+$=209.
Subtracting $E(209)$ from $E(210)$ it
yields $E_{ad}\sim$192 meV and 190 meV, respectively.
These values agree
well with those obtained with the schematic model of \cite{ya93},
which are 215 and 213 meV, respectively.

\section{Far-infrared edge modes}

In this Section we  present the results  obtained for
the far-infrared edge modes of the $N$=210 dot and compare them with
the experimental results of \cite{de90}. The method we have used
is thoroughly described in \cite{li97}. It is
based on a variational solution of the equation of motion obeyed by
the excitation operator that generates the modes, much along the
method originally proposed by Feynman to describe density excitations
of superfluid $^4$He \cite{fe72}.

The expressions to obtain the frequencies of edge
modes of different multipolarity $L$ are rather cumbersome, but only
make use of one-body densities associated with
 the ground state of the system.
In the dipole $L$=1 case, the one for which
experimental results have been unambiguously obtained, we
worked out a simple expression for the $\omega_{\pm 1}$
frequency \cite{li97}:
\begin{equation}
\omega_{\pm 1}= \sqrt{ \frac{\omega_c^2}{4} +
\frac{1}{2 N} \int d\vec{r} \,\Delta V_+(r) \,\rho(r)}
 \pm \frac{\omega_c}{2} \,\, .
\label{eq17}
\end{equation}
It is easy to further elaborate this equation and write the dipole
frequency as a function of the electric field $\cal{E}$$(r)$
created by the electrons
at the edge of the dot. We obtain:
\begin{equation}
\omega_{\pm 1}= \sqrt{ \frac{\omega_c^2}{4} +
\frac{{\cal E}(R)}{R} }
 \pm \frac{\omega_c}{2} \,\, .
\label{eq18}
\end{equation}
We have drawn $\cal{E}$$(r)$ in Fig. (\ref{fig11})
for the $B$ values corresponding to $\nu$=$\frac{2}{3}$, 1, and 4.
As it has been previously obtained  \cite{st92},
it is sharply peaked a the dot edge. The shoulders displayed by
$\cal{E}$$(r)$ near the edge reflect the complex structure of the
droplet density in that region.

We show in Fig. (\ref{fig12}) the frequencies corresponding
to the $L=\pm1,\pm2$, and $\pm3$ modes,
 together with the experimental points of
\cite{de90}. We want to stress that there are no adjustable parameters
in the calculation. From the figure, one sees that the agreement
between theoretical and experimental spectra  is good. In particular,
the $L=\pm1$ branches are well reproduced up to a high value of $B$.
The interference between the $\omega_{-3}$ and
$\omega_{+1}$ which Shikin et al.\cite{sh91} suggest it causes the level
repulsion experimentally found, is at the right place and has the
correct amplitude. In our calculation, the quadrupole branches fit
some experimental points. However, it is not yet clear whether these
points correspond to quadrupole excitations \cite{wa95}, or are the
result of a fragmentation of the dipole spectrum, as some
calculations in small dots might indicate \cite{gu91}.

In a recent work \cite{bo96},
it has been pointed out that incompressible strips at the edge of
quantum dots could be detected by far-infrared spectroscopy when the
confining potential is nonparabolic. In particular,
the low-frequency dipole branch should be especially sensitive to
the shape of the dot edge and hence, exhibits oscillations as
$B$ is varied.
Eq. (\ref{eq18}) shows that these oscillations appear if the
electric field at the edge of the dot has an oscillatory behavior with
$B$, and are absent otherwise. For example,
when the confining potential $V_+(r)$ is parabolic,
Eqs. (\ref{eq17},\ref{eq18})  yield nonoscillatory dipole
frequencies, as $\cal{E}$$/R$ is a constant. This is
easily seen from Eq. (\ref{eq17}). In particular, that constant
is $\omega_0^2 = \pi \rho_b/R$ if  the parabolic
approximation of $V_+$ in the $r/R<<1$ limit, $V_+(r)= \frac{1}{2}
\omega_0^2 r^2$, is justified.

We have analyzed the negative dispersion branch of the dipole mode
represented in Fig. (\ref{fig12}),
plotting the ratio  $\omega_-/\omega_-^{cl}$ as a function of $B$,
where
$\omega_-^{cl}$ is given by Eq. (\ref{eq18}) taking for $\cal{E}$$(R)$
its value at $B$=0.
The results are shown in Fig. (\ref{fig13}). One can see that the
oscillations in $\omega_-$ are indeed filling factor related, and as
indicated in Ref. \cite{bo96}, the minima correspond
to half filled Landau
levels (odd $\nu$), and the maxima to fully occupied Landau levels
(even $\nu$). Filling factor $\nu$=1, not attained in \cite{bo96},
is an exception as it constitutes in a sense a filled level by itself.
The dipole positive dispersion branch also presents oscillations, but
they are an order of magnitude smaller than these of the $\omega_-$
branch.

Apart from the $\nu$=1 and 2 cases, the amplitude of the oscillations
we find is smaller than in \cite{bo96}, but one should have in mind
that different confining potentials are used. It is also worth to
notice that in the present calculation, the oscillations in the
dipole
frequency are not due to the existence of integer filling
factor strips at the surface of the dot, which as we have indicated
before, do not appear here
because of the sharp jellium density we are using. Rather,
they arise due to the appearance of integer filling factor regions in
the bulk of the dot.

\section{Summary}

In this work we have used Current Density Functional Theory to describe
the ground state and multipole spectrum of a quantum dot made of 210
electrons, confined in a disk geometry and submitted to weak and
intense magnetic fields as well.
The sharp jellium confining potential does not allow the system
to  exhibit incompressible states at the edge, but its large size
permits these states to develop in the bulk. We have been able to
identify such incompressible states with filling factors from one to
ten, and many other compressible states in between.

At $B$=15.42 T, a $\nu=\frac{2}{3}$ incompressible state appears with
two-thirds of the electron number in the bulk region, and one-third in
the edge region. The occupation number of the single particle
states in the bulk region is around $2/3$, and near the edge they
rise to unity. These features have also been found in Ref.
\cite{he95} when the confining potential is supplied by a
positive jellium background, and have been interpreted there as the
formation of a composite edge in the system.

The finite temperature version of CDFT permits to solve a technical
problem when the density of s.p. states is large and it is
impracticable to solve the zero temperature KS equations with s.p.
occupation numbers 0 or 1. Interestingly, at low enough temperatures
it permits to obtain the electron density and other closely related
characteristics of highly correlated states that cannot
be described even approximately in
terms of integer occupation numbers, provided of course, these
correlations are built in the functional. This is precisely the
present case, and also that of \cite{fe95,he95}.

We predict values of the orbital and spin angular momentum that agree
well with the ones pertaining to Landau and Laughlin states. States
with integer filling factors are rather robust against changes in
$B$. In particular, the study of the total energy, ionization and
electron chemical potentials as a function of $B$ has shown that
even $\nu$, $S_z\sim0$ paramagnetic states
are especially stable, corresponding
to local minima of the total energy, and with large ionization and
electron chemical potentials. Odd $\nu$, ferromagnetic states with
large
$S_z$ values correspond to local energy maxima, apart
from the particular fully spin polarized $\nu$=1 state, and also
have comparatively small ionization and electron chemical potentials.

We have stressed the important role played by the exchange-correlation
term of the current density functional in the quantitative description
of these effects on the one hand, and the need to perform
the calculations at small temperatures to disclose fine features of
incompressible states, on the other hand.

Finally, we have studied the multipole spectrum of the dot. Good
agreement with the experimental data has been found. For the
dipole mode, we have been able to associate the oscillations with $B$
of the dipole frequencies around the classical value, with the
oscillations with $B$ of the electron electric field at the edge
of the dot. These oscillations are filling factor related and
in our case are associated to the formation of
bulk incompressible disks, and in the case of \cite{bo96}
to the formation of edge incompressible strips.

\acknowledgements

It is a pleasure to thank Nuria Barber\'an for useful discussions.
This work has been performed under grants PB95-1249 and PB95-0492
from CICYT and SAB95-0388 from DGID, Spain, and GRQ94-1022 from
Generalitat of Catalunya.

%\eject

%
%%%%%%%%%%%%%%%%%%%%%%%%%%%%%%%%%%%%%%%%%%%%%%%%%%%%%%%%%%%%%%%
%
\begin{table}
\caption{Values of $B$ and $\cal{L}$ corresponding to integer
 filling factors $\nu$=1 to 10.}
\begin{tabular}{ccc}
$\nu$ & $B$ (T) & $\cal{L}$ ($a_0^*$) \\
\tableline
1 & 10.28 & 0.82 \\
2 &  5.14 & 1.15 \\
3 &  3.43 & 1.41 \\
4 &  2.57 & 1.63 \\
5 &  2.06 & 1.82 \\
6 &  1.71 & 2.00 \\
7 &  1.47 & 2.16 \\
8 &  1.29 & 2.31 \\
9 &  1.14 & 2.45 \\
10&  1.03 & 2.58 \\
\end{tabular}
\end{table}
\begin{figure}
\caption{ Top panel: s.p. energies
as a function of orbital angular momentum $l$ at $B$=0.
The full, upright triangles represent spin up states, and the
empty, downright triangles, spin down states.
The horizontal solid line represents the electron chemical potential.
Bottom panel: The corresponding electron density as a function of
$r$, both in effective atomic units. The dotted line represents
the jellium density.}
\label{fig1}
\end{figure}
\begin{figure}
\caption{ Single particle
energies as a function of orbital angular momentum $l$
for $\nu$=1, 2, 3, 4, 6, and 8. The horizontal solid lines
represent the electron chemical potential. The full, upright
triangles
represent ($M,\uparrow$) bands, and the empty, downright triangles,
($M,\downarrow$) bands. In panel (a), only $M$=1 bands appear,
whereas in panel (f), bands with $M$=1 to 5 are visible.}
\label{fig2}
\end{figure}
\begin{figure}
\caption{Electron $\rho(r)$ (solid lines) and local
magnetization $m(r)$  (dot-dashed lines) densities
corresponding to the configurations displayed in Fig. (2).
The dashed curve in  panel (a) represents the density
of the MDD state. All densities are in effective atomic units.
Also shown are the values of ($L_z,2S_z$).}
\label{fig3}
\end{figure}
\begin{figure}
\caption{ $L_z$ (circles, left scale)
and 2$S_z$ (triangles, right scale) as a function
of $B$. The values corresponding to integer filling factors 1 to 11
are denoted by solid symbols, and the line connecting the $2S_z$
points is to guide the eye. The dashed line represents Eq. (14).}
\label{fig4}
\end{figure}
\begin{figure}
\caption{ Single particle
energies as a function of orbital angular momentum $l$
for $B$=1.9, 3, 4, 7, 9, and 11 T, respectively.
The horizontal solid
lines represent the electron chemical potential.
Same notation as in Fig. (2).}
\label{fig5}
\end{figure}
\begin{figure}
\caption{Electron $\rho(r)$ (solid lines) and local
magnetization $m(r)$     (dot-dashed lines) densities
in effective atomic units
corresponding to the configurations displayed in Fig. (5).
Also shown are the values of ($L_z, 2S_z$).}
\label{fig6}
\end{figure}
\begin{figure}
\caption{
Top panels: s.p. energies at $T$=0.1 K (dots), and at $T$=2 K
(triangles).
Bottom panels: Electron density $\rho(r)$ at $T$=0.1 K
(solid line), and at $T$=2 K (dashed line).
The left panels correspond to $B$=4 T, and the right ones
to $B$=7 T.
The electron chemical potential is indicated by a horizontal line
(solid, $T$= 0.1 K; dashed, $T$= 2 K).}
\label{fig7}
\end{figure}
\begin{figure}
\caption{ Top panels: s.p. energies (triangles, left scale), and
occupation numbers (dots, right scale),
as a function of orbital angular momentum $l$ for $B$=15.42 T.
The horizontal solid and dotted lines represent, respectively,
the electron chemical potential, and the occupation number
$\frac{2}{3}$.
Bottom panels: the corresponding electron density as a function of
$r$, both in effective atomic units. In panel (c),
the bulk density is indicated by
a dashed line, and the surface density by a dot-dashed line.
The dot-dashed line in panel (d) represents the electron density at
$T$=2 K.
Also shown are the values of $(L_z,2S_z)$. The right panels have been
obtained dropping high-$B$ contributions
to the exchange-correlation energy coming from Ref. [29].}
\label{fig8}
\end{figure}
\begin{figure}
\caption{ Top panel: Total energy  as a function of $B$.
 The values corresponding to
integer filling factors 1 to 11 are indicated by solid symbols.
The line connecting the dots is to guide the eye.
The insert shows the low-$B$ region, with the energy scale at the right.
Bottom panel: Same as top panel but as a function of $L_z$.}
\label{fig9}
\end{figure}
\begin{figure}
\caption{ Electron chemical potential (circles), and $-IP$
(triangles),
as a function of $B$. The lines connecting the points
are to guide the eye. The points corresponding to
integer filling factors 1 to 11
are indicated by solid symbols. }
\label{fig10}
\end{figure}
\begin{figure}
\caption{ Electron electric field $\cal{E}$$(r)$
in effective atomic units as a
function of $r$ for filling factors $\nu=\frac{2}{3}$, 1, and 4.}
\label{fig11}
\end{figure}
\begin{figure}
\caption{ Edge mode frequencies as a function of $B$.
At $B\sim$ 0, from top to botton the curves correspond to $L$=
$\pm3, \pm2$, and $\pm1$. The experimental dots are from
Ref. [21].}
\label{fig12}
\end{figure}
\begin{figure}
\caption{ $\omega_-/\omega_-^{cl}$ as a function of $1/B$
(lower scale) and $\nu$ (upper scale).
The solid symbols correspond to integer filling factors 1 to 11.}
\label{fig13}
\end{figure}

\begin{references}
\bibitem{note} It should be noted that in Refs. \cite{fe94,li97},
it was used an effective $\mu^*_B$ defined taking $m$ instead of
$m_e$. We have checked that the energies of the edge modes discussed
in Ref. \cite{li97} are essentially unaffected by the misuse of
$\mu^*_B$ in the Zeeman splitting.
\bibitem{note2} To avoid any misunderstanding, we remark that to
obtain the electron ionization potential we have not changed at all
the positive background, simply have subtracted one electron
and have carried out the minimization for the $N$=209,
positively charged dot.
\end{references}
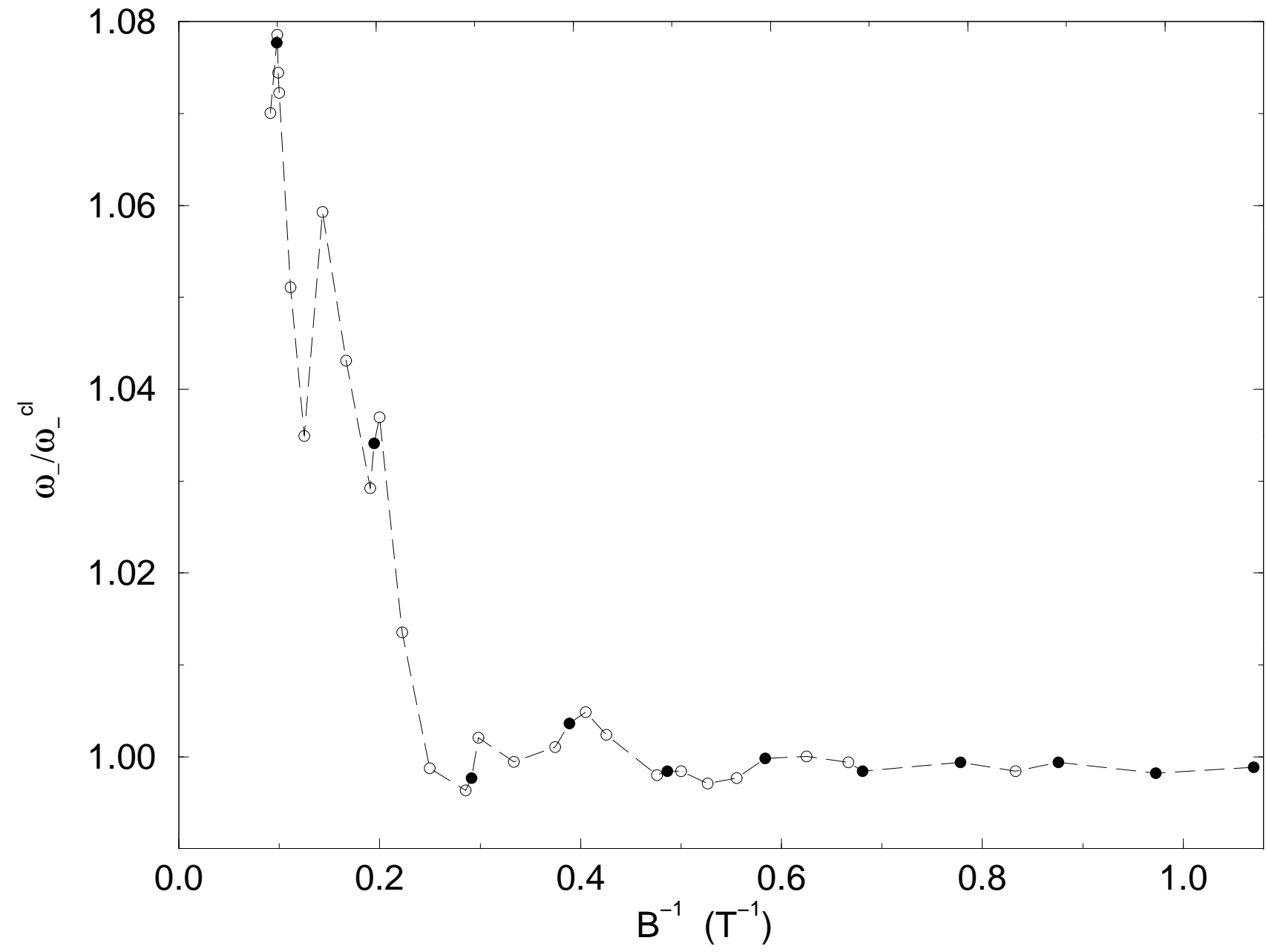
\end{document}